\documentclass[preprint,aps]{revtex4}
\begin{document}

\title{Phonon modes and heat capacity of monolayer films adsorbed in spherical pores}
\author{ R.A. Trasca$^1$, A. M. Vidales*, and  M.W. Cole$^1$}
\affiliation{$^1$Department of Physics, Pennsylvania State University and Materials Reseach Institute, University Park, PA 16802\\
* Departamento de Fisica y CONICET, Universidad Nacional de San Luis, 5700 San Luis, Argentina}

\begin{abstract}

 We examine the hydrodynamic phonon spectrum of a monolayer film
adsorbed on the wall of a spherical pore. Due to the boundary conditions,
the monolayer film exhibits a discrete phonon spectrum. The corresponding
density of states per unit frequency is thus a set of delta functions and the heat capacity
exhibits Arrhenius behavior
at low temperatures. At high temperatures, the heat capacity approaches the C $\sim T^2$
behavior of a 
two-dimensional monolayer film. Our results for the spherical surface film are compared to previous
calculations for films confined to a cylindrical surface.

\end{abstract}

\maketitle

Monolayer films have been extensively studied, both theoretically and experimentally, 
in part because they exhibit reduced dimensionality and novel phases \cite{calbi, cole1}.
For example, low pressure adsorption
on graphite usually yields a two-dimensional (2D) monolayer, whereas
adsorption in carbon nanotubes can result in quasi-one-dimensional behavior \cite{bonin,boro,cole2,hallo}.
As a result of the reduced
dimensionality, the phonon spectrum and the thermodynamic properties of such films are
altered with respect to bulk properties. The basic reason is that
the substrate imposes its boundary conditions on the adsorbate.
For example, previous calculations of  Vidales, Crespi and Cole
(VCC henceforth) show that the heat capacity of a monolayer film adsorbed on the inner wall of a 
carbon nanotube exhibits a crossover from quasi-one dimensional
to two-dimensional behavior when the thermal phonon wavelength $\lambda =\hbar s \beta$
approaches $2 \pi R$, the
circumference of the cylindrical adsorbed film (s is the sound speed and $\beta^{-1}=k_B T$) \cite{vin}.
Due to the cylindrical geometry, finite in
the radial direction and (assumed) infinite in the axial direction, the surface-parallel
phonons have a dispersion relation:
\begin{equation}
\omega_{nk}=s \sqrt{k^2+(n/R)^2}
\end{equation}
where k is the quasi-continuous wave vector (parallel to the cylinder axis) and n is the
azimuthal quantum number.

Here we consider the analogous
case of a thin film adsorbed on the inner surface of a spherical cavity.
The outline of this paper is as follows. 
We derive the in-plane phonon dispersion relation
for the spherical surface film and compare the density of states with corresponding results 
for the cylindrical and planar cases. This leads us to anticipate some of
the heat capacity results. Finally, we identify a crossover behavior in the heat
capacity and compare this with some experimental measurements.

In the derivation of the phonon dispersion relation, 
we adopt a method similar to that of the VCC paper concerned with cylindrical films
\cite{vin}. It is noted first that at low T,
the surface-normal
vibrations can be ignored since they acquire a gap at roughly the first surface-normal
excitation energy of the isolated atom ($\sim$ 60 K for $^4$He on graphite). Thus,
in a liquid monolayer film, only the surface-parallel phonons are excited at low T.
These are described by a continuous two-dimensional wave vector k for a planar film
and a one-dimensional wave vector in the cylindrical case.
Since a spherical film is confined (effectively zero dimensional),
we expect the quantum number equivalent to a wave vector k to have discrete values.
The derivation of the spectrum
follows from the scalar Helmholtz equation, describing the velocity potential $\Psi$:
\begin{equation}
\nabla^2 \Psi + k^2 \Psi =0
\end{equation}
where $k\equiv \omega /s$. This yields solutions
of the form $\Psi (\cos{\theta},\phi) = P^m _l(\cos{\theta}) \Phi(\phi)$,
where $\Phi(\phi)=e^{im\phi}, m=0,\pm 1, \pm 2,...$ and $P^m_l$ are the associated
Legendre polynomials for integers $l \ge |m|$. The boundary condition
imposed on the Legendre polynomials (to be finite when $\cos{\theta}=\pm 1$) yields the
expected discreteness of the eigenvalue $k=\sqrt{l(l+1)}/R$. The dispersion relation for
the spherical film follows as:
\begin{equation}
\omega_l=\sqrt{l(l+1)}(s/R)
\end{equation}
The corresponding density of states per unit frequency g($\omega$) is thus given by a sum
of delta functions with multiplicity given
by the m-degeneracy ($2l+1$) of each energy level:
\begin{equation}
g(\omega)=\sum_l g^{sph}_l(\omega)=\sum_l (2l+1) \delta (\omega - \omega_l)
\end{equation}

We want
to compare the spherical film results with the cylindrical and planar densities of states . For the 
cylindrical film, the number of states within the interval $d\omega$ is:
\begin{equation}
g^{cyl}(\omega) d\omega=2 \frac{L}{2\pi}dk
\end{equation}
where $L/(2\pi)$ is the number of states in the 1D reciprocal space range dk, and
2 arises from the ($\pm k$) degeneracy. Using the dispersion relation ($Eqn.1$), the cylindrical
density of states corresponding to a specific band (n) is:
\begin{equation}
g^{cyl}_n(\omega)=\frac{L}{\pi s} \frac{\Theta(\omega-s|n|/R)}{\sqrt{1-(sn/\omega R)^2}}
\end{equation}
where $\Theta(x)$ is the Heaviside step function (1 for $x \ge 0$ and 0 for $x 
< 0$). The total density of states is the sum over all bands:
\begin{equation}
g^{cyl}(\omega)=\sum_{n=-\infty}^{\infty} g^{cyl}_n(\omega)
\end{equation}
At low T, the azimuthal excitations ($n \ge 1$) are frozen out due to the gap
$E_{10}=\hbar s/R$ (see reference \cite{vin}). Thus, the dispersion relation ($Eqn.1$)
becomes $\omega=sk$ when n=0 and the cylindrical density of states becomes constant
$g_0(\omega) =\frac{L}{\pi s}$.
The planar film has also a linear dispersion relation $\omega=sk$, but k is then a 2D
wave vector. The corresponding density of states is:
\begin{equation}
g^{2D}(\omega)= \frac {A}{2\pi} \frac{\omega}{s^2} 
\end{equation}
where A is the area of the surface.

The spherical, cylindrical and planar densities of states are displayed in fig.1. In order to
get dimensionless numbers, we plot the density of states relative to the planar case for
the same area ($A=2\pi RL$ in the cylindrical case and $A=4\pi R^2$ in the spherical case).
Notice that the largest density of states at low $\omega$ corresponds to the cylindrical film, followed
by the planar film and finally, by the spherical film.This implies
that the cylindrical film will exhibit
the largest heat capacity (per unit area) at low T, followed by the planar and then, the spherical film. At 
high $\omega$, however, the number of states in an interval $d\omega$ is the
same for all types of films considered, as shown in the appendix. Therefore, the heat capacities
of the spherical, planar and cylindrical films should approach asymptotically the same
values at high T. We note that the "high T" limit in this elastic regime differs from
the very $\it high$ T limit in an experiment. That is, the law of Dulong and Petit corresponds
to a very high T $\it constant$ behavior of C. The crossover to the latter regime occurs when
$kT$ exceeds the Debye temperature. We do not consider this regime of T here.

We turn our attention to the thermodynamic properties of the spherical film. 
Using Eq.3, the energy of a phonon is:
\begin{equation}
E_l=\frac{\hbar s}{R}\sqrt{l(l+1)}
\end{equation}
There is no mode for $l=0$, and thus the spherical film possesses a gap $E_1=(\hbar s \sqrt{2})/R$. The
total energy of the phonon system is:
\begin{equation}
E_{sph}=\sum_{l=1,2...} \frac{(2l+1)E_l}{\exp(\beta E_l)-1}
\end{equation} 
where 2l+1 is the degeneracy of the energy levels. In the high
T limit, the sum can be replaced by an integral, which can be evaluated analytically
(because $1\ll l$ for the predominant modes):
\begin{equation}
E_{sph} \sim \frac{\hbar s}{R} \int_0 ^\infty \frac{2l^2 dl}{exp(\beta \hbar sl/R)-1}=\frac {A \zeta (3)}{\pi \beta^3 \hbar^2 s^2}
\end{equation}
where $A=4 \pi R^2$ is the film area and $\zeta$ is the Riemann zeta function.
As in the case of the cylindrical film, the high
T heat capacity per unit area of the spherical film is the same as that of a planar film:
\begin{equation}
\frac{C_{2D}(T)}{k_B A}=\frac{3 \zeta(3)}{\pi \beta^2 \hbar^2 s^2}
\end{equation}
For simplicity, we define a reduced temperature:
\begin{equation} 
t=R/(\beta \hbar s)
\end{equation}
and the dimensionless heat capacity takes the form:
\begin{equation}
C_{sph}/k_B=t^{-2} \sum_{l=0,1,...} \frac {l(l+1)(2l+1)e^{\sqrt{l(l+1)}/t}}{(e^{\sqrt{l(l+1)}/t}-1)^2}
\end{equation}
As in the case of the cylindrical
film \cite{vin}, the heat capacity is seen to be a universal function of the reduced temperature (the right side
of Eq.13).
$Fig. 2$ shows the dependence on T of the spherical and cylindrical heat
capacities, compared to the planar heat capacity ($C/C_{planar}$).
Notice that the planar film is the high T limit for both the spherical and cylindrical
films, as expected. However, the heat capacity of the cylindrical film approaches it from above,
whereas the heat capacity of the spherical film approaches it from below. This behavior was
expected from the previous discussion of the densities of states. We note
that the asymptotic approach to the planar limit occurs at a lower T in the cylindrical
case than in the spherical case. This can be explained in at least two ways. One is that
a cylinder (having one infinite radius of curvature) is closer to a plane than a sphere
is. The other way is that the density of states (Fig.1) of the cylinder is closer to that
of a planar surface than is the density of states for the sphere.
In analogy to the crossover behavior of the cylindrical film
\cite{vin}, one may speak of a
crossover temperature ($T_{cross}$) from a zero-dimensional system to a two-dimensional system,
defined as that temperature where the
heat capacity is about half of that of a planar film. In the present case, from fig.2,
$k_B T_{cross} \sim \frac {\hbar s}{4R}$.
 At low T, the spherical heat capacity has an activated behavior: 
\begin{equation}
C_{sph}/k_B=6 t^{-2} exp(-\sqrt{2}/t)
\end{equation}

We do not know of any experiments related to adsorption in spherical pores, but
the irregular confined pores of a disordered material may conceivably be approximated as
spherical pores. Crowell and Reppy 
have investigated $^4He$ films adsorbed in two porous glasses, Vycor and aerogel,
using high-precision torsional oscillator and calorimetry techniques \cite{gap1,gap2,gap3}. 
At some critical coverage, the $^4$He film undergoes a transition from nonsuperfluid to
superfluid behavior. In the nonsuperfluid regime, the heat capacity exhibits a linear
dependence at large T, but C/T drops rapidly at low T, so a crossover
temperature between these two behaviors is defined and calculated to be about 100 $mK$
for the lowest coverage.
In the context of our calculations, a working hypothesis is that the pore geometry is
like an American football, i.e. an ellipse of revolution. This could show the behavior of a 
cylinder ($C\sim T$) at modest T, but exhibit a gap at low T, as in the case of a spherical
geometry. We may estimate the heat capacity at low T with our spherical pore model, and
use the cylindrical pore for higher T. For a pore
diameter of about 70 $\mathring{A}$, as in the case of Vycor pores, we assume that the $^4He$
atoms form a spherical film of about 60 $\mathring{A}$ diameter, so there are about $10^3$
He atoms in a pore. Then the heat capacity per mole is of the order of mJ/K, the same
order of magnitude as the experimental heat capacity of the nonsuperfluid $^4He$ film.
However, the calculated heat capacity
exhibits the characteristic activated behavior seen in $Eq.15$
and does not follow the general trend of the experimental measurements, which looks more like
a power law behavior. Also, the heat capacity experimental values at higher T can be
obtained with our cylindrical pore model but only with an unreasonably large speed of sound.
However, the crossover temperature given by our spherical model is of the
order of 150 mK, somewhat comparable to the experimental crossover of 100 mK. Thus, it may
happen that the geometry of the Vycor pores produces boundary conditions for the film, 
whose effect changes with increasing temperature, leading
to a crossover behavior.

In conclusion, the substrate geometry imposes constraints on the phonons of
monolayer films adsorbed within a spherical pore. The first effect of the confinement is the discreteness of the
phonon spectrum. Because of a low $\omega$ gap in the spectrum,
the specific heat exhibits an essential singularity at T=0, whereas
at high T, it asymptotes to that of a planar film ($C\sim T^2$). A somewhat different crossover behavior occurs in the case of
a cylindrical film, which has quasi-one-dimensional behavior at low T ($C\sim T$) and
planar (2D) behavior at high T. This pair of solutions may explain some of the heat capacity
features of monolayers adsorbed in disordered porous materials, whose various pores have
different geometries and sizes.

We are grateful to the National Science Foundation, which has supported this research. We thank
Paul Crowell and Moses Chan for helpful communications.

\appendix{APPENDIX}
\maketitle

This paper states that the density of states at high $\omega$ is the same for all three
problems discussed here. We will give now a proof of this statement.
Let us start with the case of a spherical film, and choose an interval $d\omega$ between
$\omega_1=\sqrt{l(l+1)}(s/R)$ and $\omega_2=\sqrt{(l+1)(l+2)}(s/R)$, corresponding
to consecutive values of l when $l\gg 1$. The number of states in this
interval is:
\begin{equation}
\int_{\omega_1}^{\omega_2} g^{sph}(\omega) d\omega = (2l+1) \int_{\omega_1}^{\omega_2} \delta (\omega-\omega_l)=2l+1
\end{equation}
Thus, in the limit of high $\omega$, and correspondingly high l, the number of states
in an interval $d\omega$ is of the order 2l. 

For the planar film, we consider the same interval $d\omega$, and the number of states
in this interval is:
\begin{equation}
\int_{\omega_1}^{\omega_2} \frac {A}{2 \pi s^2} \omega d\omega =2(l+1)
\end{equation}
where the area of the planar film equals the area of the
spherical film $A=4\pi R^2$.
For high l, the number of states of the planar film in this interval $d\omega$ is therefore of the order of
$2l$, equal to that
of the spherical film. 

The last case is that of a cylindrical film. The number of states corresponding to the
interval $d\omega$ is:
\begin{eqnarray*}
\int_{\omega_1}^{\omega_2} \frac {L}{\pi s} (\sum_{n=-l}^l \frac{\Theta(\omega-sn/R)}{\sqrt{1-(sn/\omega R)^2}}) d\omega &=& \frac{L}{\pi s} \sum_{n=-l}^l \int_{\omega_1}^{\omega_2}\frac{d\omega}{\sqrt{1-sn/\omega R)^2}} \nonumber\\
                                                                                                                         &=& \frac{L}{\pi R}\sum_{n=-l}^l (\sqrt{(l+1)(l+2)-n^2}-\sqrt{l(l+1)-n^2}) 
\end{eqnarray*}
First, in order to compare the cylindrical and spherical films, we take their areas to be
the same; thus $2\pi R L=4 \pi R^2$ or $L=2R$. Then, if we define
$\sqrt{l(l+1)-n^2}$ as $f_n (l)$, a function of variable l, the result of
integration can be written as $f_n(l+1)-f_n(l)$, which (for large l) is the derivative
$f_n ' (l)$. In the
limit of large l, $f_n(l)\sim \sqrt{l^2-n^2}$, whose derivative is $l/\sqrt{l^2-n^2}$.
Converting the sum to an integral the number of states becomes:
\begin{eqnarray*}
N_{states} &=& \frac{2}{\pi} \sum_{n=-l}^l \frac{l}{\sqrt{l^2-n^2}} \nonumber\\
		&\simeq& \frac{2}{\pi} l \int_{-l}^l \frac {dn}{\sqrt{l^2-n^2}} \nonumber\\
		&=& 2l
\end{eqnarray*}

 Therefore, at high frequency, the integrated number of states in the
interval $d\omega$ is the same in all three cases.

\newpage

pictures

1. The density of states of the cylindrical film
(dashed curve) relative to that of planar film (dotted line) as a function of 
the reduced frequency $\omega R/s$.The full vertical lines indicate the delta functions
$g_{sph}(\omega)$ for the spherical film.

2. The heat capacity of a cylindrical (dashed) and spherical (full) film relative
to that of a planar film, as a function of the reduced temperature defined in $Eq.13$.

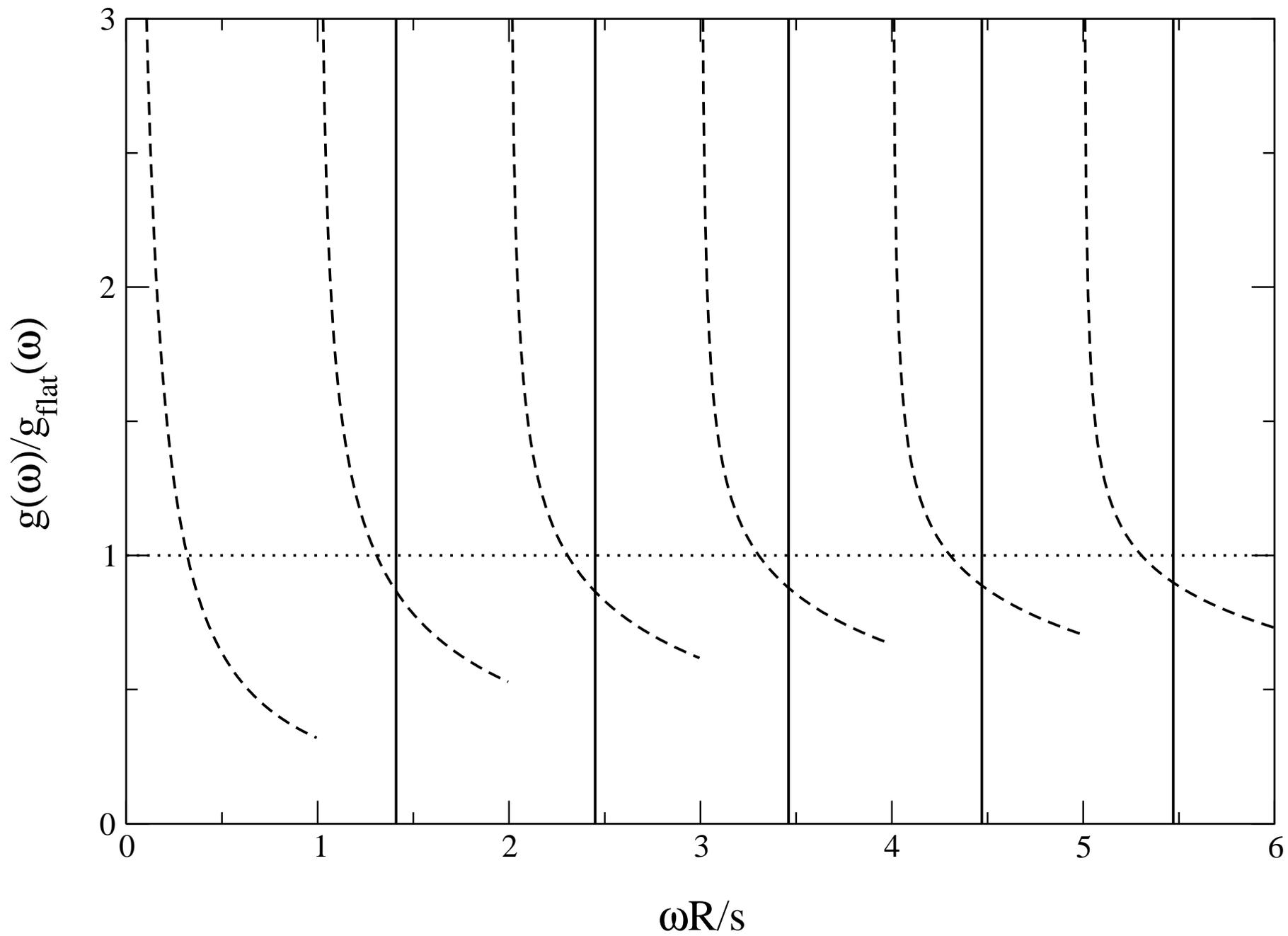
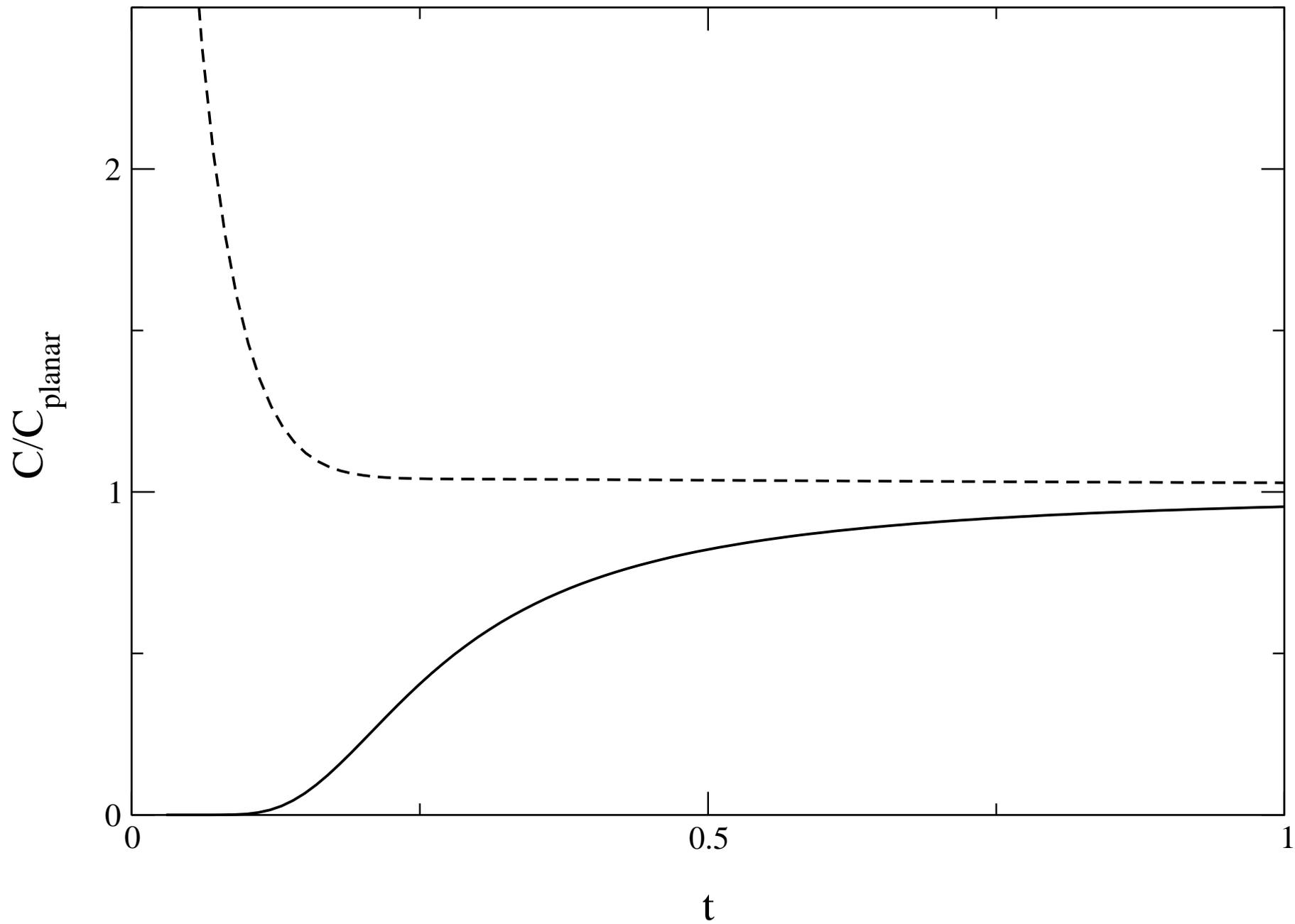
\end{document}